\documentclass[10pt]{iopart}
\usepackage{graphicx}
\usepackage{siunitx}
\usepackage{cite}

\begin{document}
\title[Highly selective dry etching of GaP in the presence of Al$_\textrm{x}$Ga$_{1-\textrm{x}}$P]{Highly selective dry etching of GaP in the presence of Al$_\textrm{x}$Ga$_{1-\textrm{x}}$P with a SiCl$_4$/SF$_6$ plasma}

\author{Simon H\"onl, Herwig Hahn, Yannick Baumgartner, Lukas Czornomaz and Paul Seidler}

\address{IBM Research {--} Zurich, S\"aumerstrasse 4, CH-8803 R\"uschlikon, Switzerland}
\eads{\mailto{hon@zurich.ibm.com} and \mailto{pfs@zurich.ibm.com}}

\begin{abstract}
	We present an inductively coupled-plasma reactive-ion etching process that simultaneously provides both a high etch rate and unprecedented selectivity for gallium phosphide (GaP) in the presence of aluminum gallium phosphide (Al$_\textrm{x}$Ga$_{1-\textrm{x}}$P).  Utilizing mixtures of silicon tetrachloride (SiCl$_4$) and sulfur hexafluoride (SF$_6$), selectivities exceeding 2700:1 are achieved at GaP etch rates above 3000 nm/min. A design of experiments has been employed to investigate the influence of the inductively coupled-plasma power, the chamber pressure, the DC bias and the ratio of SiCl$_4$ to SF$_6$. The process enables the use of thin Al$_\textrm{x}$Ga$_{1-\textrm{x}}$P stop layers even at aluminum contents of a few percent.
\end{abstract}


\maketitle



\section{Introduction}
III-V materials are well established in the semiconductor industry for applications ranging from RF amplifiers in cellular communication devices to light emitting and laser diodes \cite{kneissl2010advances,chang2012light,kish1994very,ponce1997nitride} to multi-junction concentrator solar cells \cite{yamaguchi2005multi, bett1999iii}. The integration of such devices onto silicon substrates is an emerging trend, offering the opportunity to combine, for example, low-power opto-electronic devices \cite{fang2006electrically,dong2014novel,mathine1997integration} and high-electron-mobility transistors \cite{kazior2011high} based on III-V compound semiconductors with complex silicon microelectronic circuits. Numerous techniques have been developed for the microfabrication of such III-V devices \cite{pearton1994reactive}, among which processes for selective etching are essential.\\ 
\indent Inductively coupled-plasma reactive-ion etching (ICP-RIE) is a particularly attractive method to achieve selectivity because the separation of plasma generation and ion bombardment offers additional flexibility for optimizing the etching of one material preferentially over another \cite{shearn2010advanced}. ICP-RIE processes have been developed for compounds such as GaAs, AlGaAs, InGaAs, InP and GaN \cite{lee2000advanced,karouta1999chemical,volatier2010extremely,maeda1999inductively,maeda1999inductively2,hahn2012influence}. In contrast, GaP is a largely unexplored material for integrated nanophotonic and opto-electronic applications, despite its attractive combination of large refractive index and large electronic bandgap, and fabrication processes for this material are in general far less advanced. In particular, selective etching of GaP versus Al$_\textrm{x}$Ga$_{1-\textrm{x}}$P has hardly been investigated.  Although wet etch processes for the selective removal of Al$_\textrm{x}$Ga$_{1-\textrm{x}}$P in the presence of GaP exist \cite{lee1996wet}, there are no reports of the opposite selectivity by wet etching, and only one such dry etching process has been previously investigated \cite{epple2002dry}.  The etch rate and selectivity of the latter process are too low for the practical removal of large amounts (many tens of microns) of GaP or for stopping on a thin layer (a few hundred nanometers) of Al$_\textrm{x}$Ga$_{1-\textrm{x}}$P, as may be needed for process flows involving wafer bonding, for example.\\
\indent To address this shortcoming, we present here an ICP-RIE process based on SiCl$_4$ and SF$_6$ that etches GaP over Al$_\textrm{x}$Ga$_{1-\textrm{x}}$P with unprecedented selectivity while maintaining an extremely high GaP etch rate. The process enables the use of thin Al$_\textrm{x}$Ga$_{1-\textrm{x}}$P etch stop layers with aluminum content as low as a few percent, providing much greater process control and enabling new types of GaP-based devices \cite{schneider2017optomechanics}. We employ the design-of-experiments method \cite{myers2016response,ilzarbe2008practical,fisher1937design} to investigate the influence of pressure, ICP power, DC bias and gas composition on the etch rate of GaP and the selectivity of etching GaP with respect to two Al$_\textrm{x}$Ga$_{1-\textrm{x}}$P compositions. 

\section{Experimental Methods}
\subsection{Sample preparation}
\begin{figure}[h]
	\centering
	\includegraphics[width = 8cm]{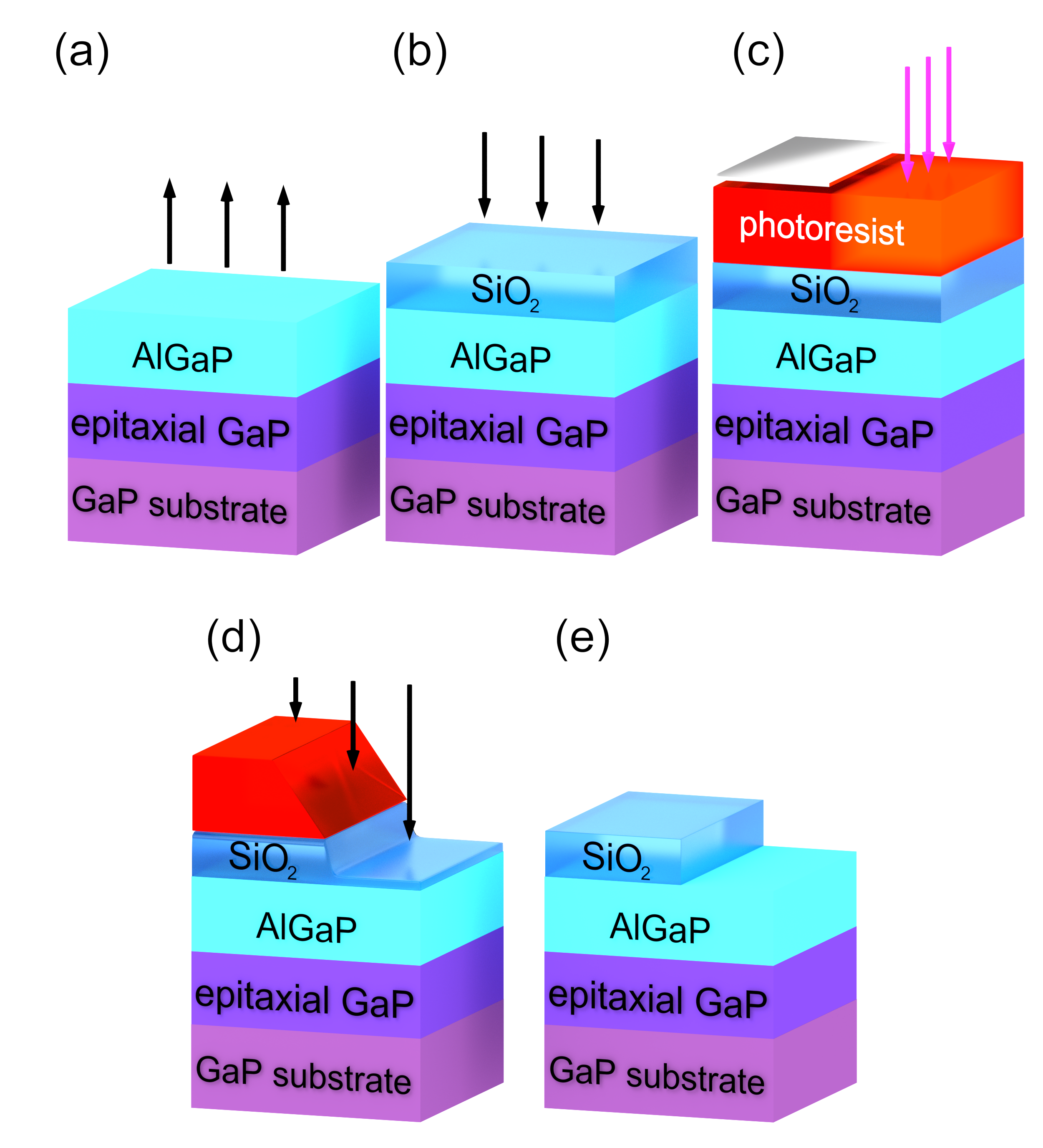}
	\caption{Process flow for the fabrication of Al$_\textrm{x}$Ga$_{1-\textrm{x}}$P samples: (a) epitaxial growth of GaP and Al$_\textrm{x}$Ga$_{1-\textrm{x}}$P by MOCVD; (b) deposition of SiO$_2$ by PECVD; (c) spin-coating and exposure of photoresist; (d) pattern transfer into SiO$_2$ by RIE; (e) removal of photoresist.}
	\label{fig1}
\end{figure}

Al$_\textrm{x}$Ga$_{1-\textrm{x}}$P samples were epitaxially grown by metalorganic chemical vapor deposition (MOCVD) (Veeco P125) from trimethylgallium (TMGa), trimethylaluminum (TMAl) and tertiarybutylphosphine (TBP) on undoped, [100]-oriented quarters of a 2-inch GaP wafer with a nominal thickness of \SI{400}{\micro\meter}. Chips diced from the same ‘epi-ready’ substrates were used as samples for the etching experiments on GaP itself. The growth temperature in the MOCVD system was \SI{650}{\degree C} at the susceptor as measured with an optical pyrometer. Prior to the growth of the Al$_\textrm{x}$Ga$_{1-\textrm{x}}$P layers, a \SI{100}{nm}-thick homoepitaxial layer of GaP was grown at a V-to-III molar-flow ratio of 10:1. Subsequently, Al$_\textrm{x}$Ga$_{1-\textrm{x}}$P layers with a target thickness of \SI{100}{\nano\meter} were grown at a V-to-III molar-flow ratio of 6:1 with various proportions of TMGa and TMAl. The resulting film stoichiometries were determined by X-ray diffraction to have an aluminum content of x = 0.033 and x = 0.097 and are referenced below as x = 0.03 and x = 0.10, respectively. \\
\indent A SiO$_2$ hard mask was employed for the etching experiments on both the GaP and Al$_\textrm{x}$Ga$_{1-\textrm{x}}$P samples.  Starting with a quarter GaP wafer and a quarter wafer of each Al$_\textrm{x}$Ga$_{1-\textrm{x}}$P composition, \SI{100}{nm} of SiO$_2$ were deposited by plasma-enhanced chemical vapor deposition (PECVD) in an Oxford Plasmalab 100 system with SiH$_4$ and N$_2$O as precursors. The SiO$_2$ was then photolithographically patterned using the positive resist AZ6612 (Microchemicals GmbH) and a CF$_4$/Ar RIE process in an Oxford NGP 80 system. The photoresist was subsequently removed using acetone, isopropanol, and an oxygen plasma (GIGAbatch 310 M from PVA TePla), which should also eliminate non-volatile etch residues originating from the CF$_4$-containing RIE process. Finally, the GaP and Al$_\textrm{x}$Ga$_{1-\textrm{x}}$P quarter wafers were diced into 3.5-mm $\times$ 3.5-mm chips. The complete process flow is schematically illustrated in \fref{fig1}.\\
\subsection{Etching}
For the development of the selective ICP-RIE process with SiCl$_4$ and SF$_6$, the samples were etched for \SI{300}{s} in an Oxford Instruments PlasmaPro 100 ICP system. The samples were placed side by side at the center of a 4-inch Si carrier wafer with no additional thermal conduction promoter such as oil, wax, or grease. The temperature of the carrier wafer was held at \SI{30}{\degreeCelsius} with a helium backing flow. The other process parameters were determined by the design of experiments (see below). \\
\indent Following etching, the surface profile of the samples was measured with a Bruker DekTak XT profilometer, where the measured step height included the residual SiO$_2$ hard mask. The hard mask was then stripped with a 90-s dip in buffered hydrofluoric acid (BHF), after which the etch profile was remeasured to determine both the etch rate and how much of the hard mask had been removed during the ICP-RIE process. The etch rate of GaP and Al$_\textrm{x}$Ga$_{1-\textrm{x}}$P (x = 0.03 and x = 0.10) in BHF is negligible (measured to be $< \SI{0.3}{nm\per min}$ for x = 0.10 by atomic force microscopy in a separate control experiment using an organic resist for the mask). If it was found that the SiO$_2$ layer had been completely consumed during dry etching, the experiment was repeated for an etch duration of \SI{120}{s} . To account for differences in the etch rate across the chip, the profile was measured at three positions from the edge towards the center in steps of \SI{600}{\micro\meter} and the results averaged.

\subsection{Design of Experiments}
Given the large parameter space of an ICP-RIE process \cite{pearton1994reactive}, it is obviously time-consuming and laborious to analyze the process by investigating changes of one parameter at a time. To obtain a general understanding of the etch process more efficiently, i.e. without performing experiments for every possible parameter combination, we used a design-of-experiments (DOE) approach \cite{myers2016response}, probing a limited selection of points distributed over the parameter space. Specifically, a fractional factorial design comprising 24 experiments was created with the commercial statistical software JMP\textsuperscript{\textregistered} (SAS Institute) \cite{jmp}. We investigated the influence of the ICP power ($P_\textrm{ICP}$), the chamber pressure ($p_\textrm{C}$), the DC bias voltage ($V_\textrm{DC}$), and the SiCl$_4$-to-SF$_6$ ratio expressed as the fraction of SiCl$_4$ ($f_{\textrm{SiCl}_4}$) on the etch rate of GaP and the selectivity between GaP and Al$_\textrm{x}$Ga$_{1-\textrm{x}}$P. The parameters of the individual experiments performed as specified by JMP for an I-optimality criterion \cite{myers2016response} are displayed in \tref{tab1}. An I-optimal design was possible because preliminary experiments gave us confidence in having already obtained a general understanding of the effect of the process conditions, and it provides for better prediction than e.g. the D-optimal design \cite{myers2016response}. Values for the ICP power, pressure, and DC bias were chosen by the software from a continuous range. The chamber pressure was varied between \SI{5}{mTorr} and \SI{60}{mTorr} and the ICP power between \SI{25}{W} and \SI{400}{W}. The DC bias was set to values in the range of \SI{50}{V} to \SI{250}{V} by adjusting the RF power for each experiment individually.  For $f_{\textrm{SiCl}_4}$, six explicit values were chosen according to preliminary experiments, keeping the combined total flow of SiCl$_4$ and SF$_6$ constant at \SI{20}{sccm}.

\begin{table*}	
	\caption{Experimental parameters and resulting etch rates. $\rho_i$ denotes the etch rate for the various materials with $i$ indicating the aluminum content in Al$_\textrm{x}$Ga$_{1-\textrm{x}}$P ($i$ = 0 represents GaP). Becuase the signal-to-noise ratio of step heights less than \SI{10}{nm} is low, the etch rates below \SI{2}{nm\per min} generally have a large relative error. Values noted as 0 correspond to no resolvable steps observed with the profilometer. Etch rates that were excluded from the DOE fit due to residue deposition or micomasking are marked with an asterisk, in which case the given value for the etch rate generally represents a lower limit. }

	\begin{indented}
		\item[]\begin{tabular}{@{}llllllll}
			\br
			$\textrm{No.}$ & \begin{tabular}{@{}l}
				$f_{\textrm{SiCl}_4}$ \\ $[\%~ \textrm{SiCl}_4]$\end{tabular}  & \begin{tabular}{@{}l}
			$p_\textrm{C}$\\ $ [\SI{}{mTorr}]$ \end{tabular}   & \begin{tabular}{@{}l}
			$V_\textrm{DC}$\\ $[\SI{}{V}]$ \end{tabular}   & \begin{tabular}{@{}l}
			$P_\textrm{ICP}$\\ $[\SI{}{W}]$ \end{tabular}   & \begin{tabular}{@{}l}
			$\rho_0$\\ $[\SI{}{nm}]$ \end{tabular}  & \begin{tabular}{@{}l}
			$\rho_{0.03}$\\ $[\SI{}{nm}]$ \end{tabular}& \begin{tabular}{@{}l}
			$\rho_{0.10}$\\ $[\SI{}{nm}]$ \end{tabular} \\ \mr
			$1$ & $37.5$ & $51.8$ & $97$ & $25$ & $2.39$ & $4.06$ & $4.51$ \\
			$2$ & $75$ & $5$ & $60$ & $64$ & $114$ & $33.7$ & $10.6$ \\ 
			$3$ & $50$ & $49.1$ & $255$ & $347$ & $3394$ & $1.5$ & $1.25$ \\ 
			$4$ & $50$ & $16.1$ & $207$ & $25$ & $1209$ & $53.3^*$ & $11.9$ \\
			$5$ & $50$ & $35.3$ & $166$ & $194$ & $2637$ & $6.6$ & $3.59$ \\
			$6$ & $25$ & $49.1$ & $77$ & $403$ & $2.4$ & $4.35$ & $4.34$ \\
			$7$ & $37.5$ & $60.1$ & $211$ & $403$ & $8.26$ & $11.1$ & $10.1$ \\
			$8$ & $66.7$ & $51.7$ & $122$ & $308$ & $2527$ & $22.4$ & $12.9$ \\
			$9$ & $25$ & $5$ & $202$ & $308$ & $398^*$ & $122^*$ & $111^*$ \\
			$10$ & $100$ & $60.1$ & $108$ & $25$ & $24.9$ & $19.9$ & $17.5$ \\
			$11$ & $37.5$ & $5.1$ & $250$ & $103$ & $346$ & $327$ & $229$ \\
			$12$ & $25$ & $60.1$ & $260$ & $45$ & $11.7$ & $9.86$ & $10.4$ \\
			$13$ & $37.5$ & $46.2$ & $183$ & $103$ & $6.3$ & $4.15$ & $4.45$ \\
			$14$ & $50$ & $60.1$ & $51$ & $140$ & $503^*$ & $11.5$ & $5.57$ \\
			$15$ & $66.7$ & $18.8$ & $232$ & $403$ & $1129$ & $98.6^*$ & $44.8^*$ \\
			$16$ & $0$ & $32.6$ & $230$ & $271$ & $1.84$ & $2.08$ & $1.89$ \\
			$17$ & $25$ & $21.5$ & $100$ & $121$ & $2.73$ & $2.53$ & $1.89$ \\
			$18$ & $37.5$ & $13.3$ & $50$ & $347$ & $4.45$ & $3.79$ & $3.34$ \\
			$19$ & $37.5$ & $18.7$ & $100$ & $289$ & $5.6$ & $5.46$ & $3.91$ \\
			$20$ & $0$ & $5$ & $69$ & $25$ & $0$ & $0$ & $0$ \\
			$21$ & $100$ & $5$ & $209$ & $233$ & $27.5$ & $27.3$ & $26.2$ \\
			$22$ & $100$ & $41.3$ & $49$ & $403$ & $6.06$ & $3.09$ & $1.7$ \\
			$23$ & $50$ & $5$ & $100$ & $403$ & $216$ & $48.8^*$ & $17.4$ \\
			$24$ & $66.7$ & $49.1$ & $233$ & $103$ & $3008$ & $133^*$ & $57.4^*$ \\ \br
		\end{tabular}
	\end{indented}
	\label{tab1}
\end{table*}

\section{Results and Analysis}
The etch rates obtained from the 24 experiments comprising the DOE are presented in \tref{tab1}. The results were analyzed separately with respect to two responses:  the GaP etch rate and the selectivity between GaP and Al$_\textrm{x}$Ga$_{1-\textrm{x}}$P. Because the observed etch rates span several orders of magnitude and the error is expected to increase with both etch rate and selectivity, the least-squares regression analyses with JMP were performed on the common logarithm of the response (either etch rate or selectivity). Much better fits to the data were thus achieved; a systematic skew of the distribution of residuals as a function of the response was observed with a linear scale but not with a logarithmic scale.\\
\indent To find an accurate model of minimal complexity, we started in each case by describing the desired response as an over-specified polynomial function of the process parameters, including cross terms. The coefficient of determination R$^2$, the adjusted R$^2$, and the probability value $p$ for both individual terms and the model as a whole were calculated. Terms with low statistical significance as gauged by a high $p$-value were sequentially removed until the highest remaining individual $p$-value was below 0.05 and removal of any more terms would lead to a substantial drop in the adjusted R$^2$. The global maximum of the response within the explored parameter range was also determined as well as the variation of the response in the vicinity of the global maximum as a function of each process parameter individually.

\subsection{GaP etch rate}
\begin{figure}[h]
	\centering
	\includegraphics[width = 8cm]{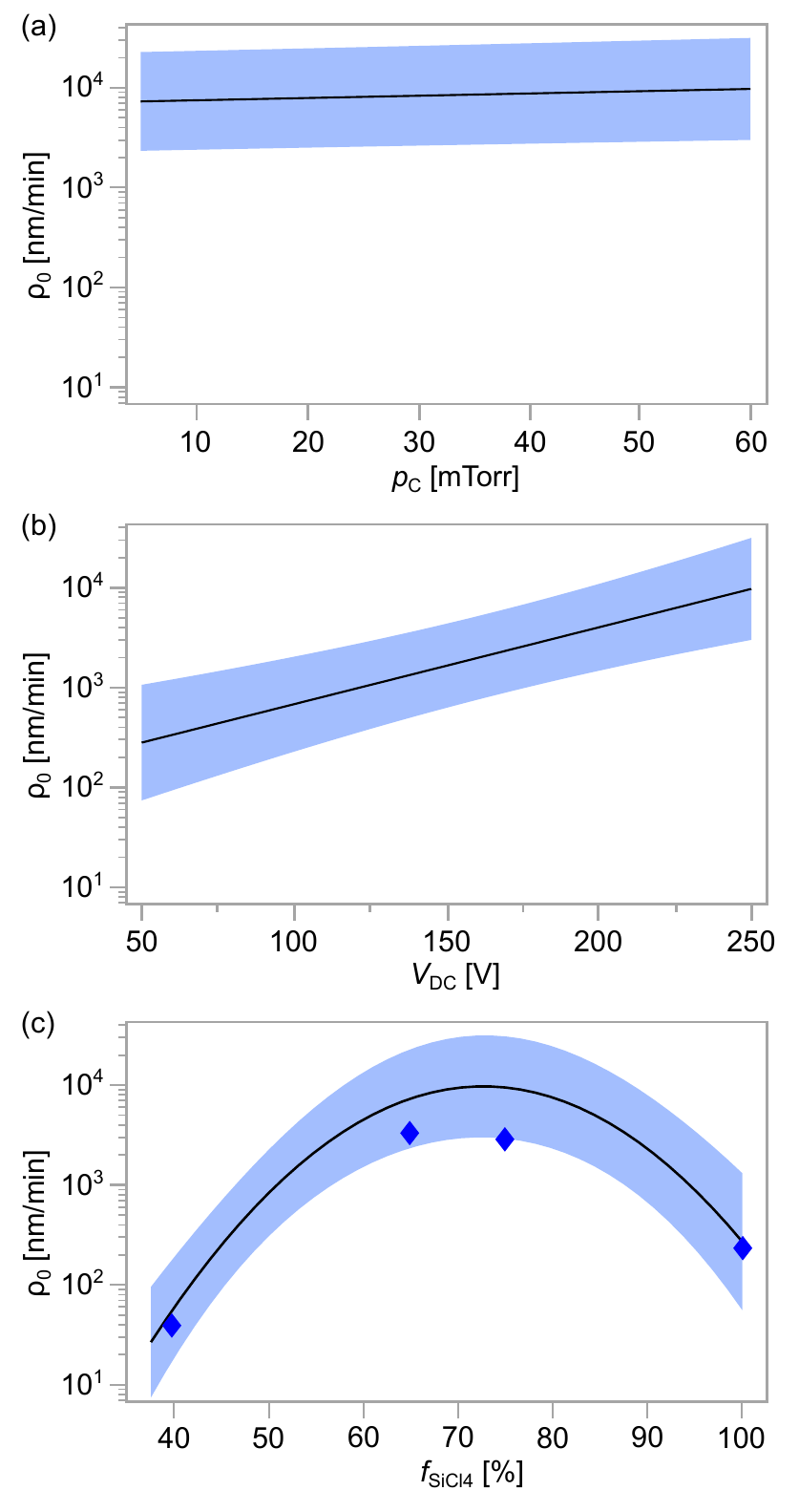}
	\caption{GaP etch rate as predicted by the model as a function of (a) $p_\textrm{C}$, (b) $V_\textrm{DC}$, and (c) $f_{\textrm{SiCl}_4}$, individually. In each panel, the other free parameters are fixed at $p_\textrm{C} = \SI{60}{mTorr}$, $V_\textrm{DC} = \SI{250}{V}$, and $f_{\textrm{SiCl}_4} = 73\%$, respectively. The blue shaded regions define 95\% confidence intervals. The dark blue diamonds in (c) indicate the measured etch rate for additional etch experiments at the indicated SiCl$_4$ content.}
	\label{fig2}
\end{figure}

\begin{figure}[h]
	\centering
	\includegraphics[width = 8cm]{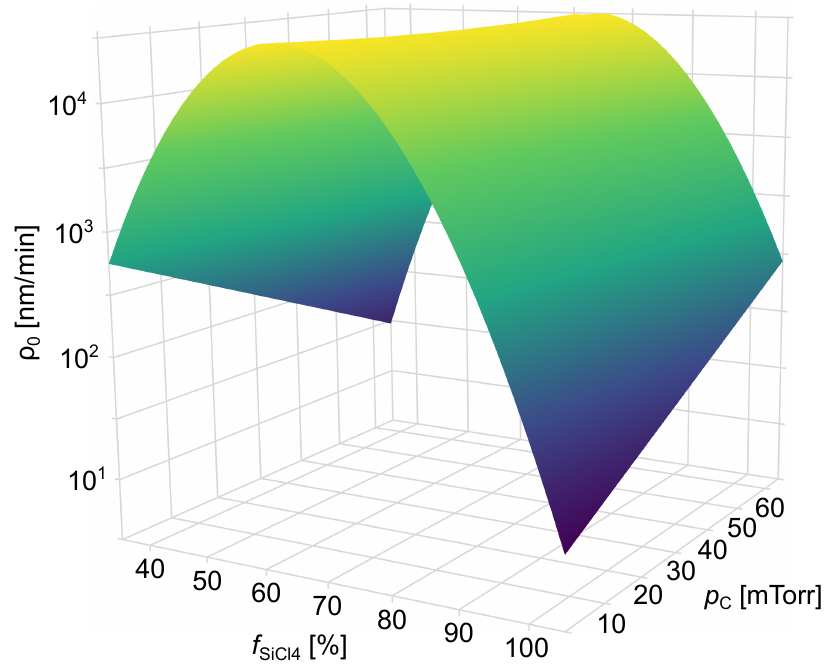}
	\caption{Predicted GaP etch rate plotted versus chamber pressure ($p_\textrm{C}$) and fraction of SiCl$_4$ ($f_{\textrm{SiCl}_4}$).}
	\label{fig3}
\end{figure}

\begin{figure}[h]
	\centering
	\includegraphics[width = 8cm]{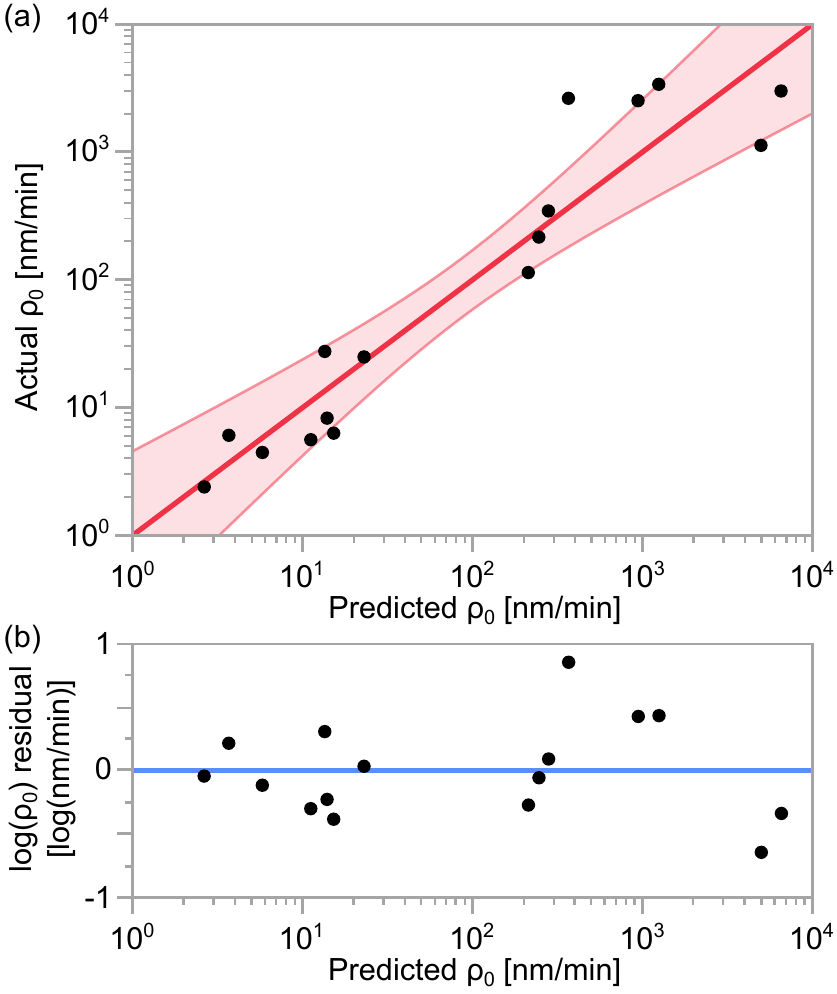}
	\caption{(a) Actual-by-predicted plot of the GaP etch rate $\rho_0$ with the 95\% confidence interval shaded in red, and (b) plot of residuals of $\log(\rho_0)$ as a function of predicted $\rho_0$. In the absence of errors, the black dots indicating the individual experiments would lie on the straight lines.}
	\label{fig4}
\end{figure}

We first present a model for the GaP etch rate alone. (Selectivity is treated with a separate model to account for the additional mechanisms in effect in the etching of Al$_\textrm{x}$Ga$_{1-\textrm{x}}$P that do not pertain to GaP.)  We restrict the model to processes for which the fraction of SiCl$_4$ is within the range of 37.5\% to 100\% because the abrupt change from the low etch rates observed at lower fractions of SiCl$_4$ would require a high-degree polynomial with more terms than can be justified with the size of the data set.  In other words, we model a region of the parameter space for which GaP is etched at an appreciable rate.  In addition, experiment 14 was excluded, as residues formed on this sample during the etch process leading to micromasking effects, so the etch rate could not be unambiguously determined. \\
\indent After thorough investigation of various alternative polynomial functions of the process parameters, we consistently found that only the three terms listed in \tref{tab2} are necessary to provide a suitable model. The calculated dependence of the GaP etch rate on each individual process parameter with the others fixed at the value that gives the maximum response is displayed in \fref{fig2}. The ICP power dependence is not shown because this parameter is not statistically significant over the parameter range investigated, a result that suggests that the plasma is already saturated at the lowest value of \SI{25}{W} with those species involved in the etching of GaP. \\
\begin{table}
	\caption{Model terms and their corresponding probability value \textit{p} for the etch rate of GaP. }
	\begin{indented}
		\item[]\begin{tabular}{@{}ll}
			\br
			$\textrm{Term}$ & $p$ \\ \mr
			$f_{\textrm{SiCl}_4}^2$ & $0.00000$  \\ 
			$V_\textrm{DC}$ &$ 0.00041$ \\ 
			$p_\textrm{C}\cdot f_{\textrm{SiCl}_4}$  & $0.01668$ \\ \br
		\end{tabular}
	\end{indented}
	\label{tab2}
\end{table}
\indent While the chamber pressure alone is also not significant, it does play a role in conjunction with the fraction of SiCl$_4$.  Indeed, it is not surprising that the product of pressure and SiCl$_4$ fraction ($p_\textrm{C}\cdot f_{\textrm{SiCl}_4}$) appears in the model, as this term is related not only to the relative but also the absolute concentration of the etchants. The effect of $p_\textrm{C}$ is primarily to shift the value of $f_{\textrm{SiCl}_4}$ for which the maximum etch rate is achieved, not to increase the etch rate, as is evident from \fref{fig2}(a), in which the modest monotonic increase in etch rate with pressure stays within the 95\% confidence interval. The model predicts that for high SiCl$_4$ content a higher pressure is preferable, whereas at high SF$_6$ content a lower pressure yields a higher etch rate. This behavior is evident in the response surface plot in \fref{fig3}.\\ 
\indent We interpret this finding as indicating that species originating from SiCl$_4$ rather than SF$_6$ dominate chemical reaction with GaP.  Specifically, etching by ion bombardment of the surface is more important when the amount of chlorine-containing species is low, as this physical process should benefit from lower pressure. If the content of chlorine-containing species in the plasma is high, chemical reactions with the surface play a greater role and should be enhanced at high ion density, i.e. at high pressure.\\
\indent Of considerably more statistical significance is the DC bias.  The monotonic increase in etch rate over one and a half orders of magnitude as the DC bias increases from \SI{50}{V} to \SI{250 }{V} (\fref{fig2}(b)) is attributed to the change in kinetic energy of the ions bombarding the sample and the consequent enhancement of the physical component of etching. We note that the calculated optimum value for $V_\textrm{DC}$ corresponds to one end of the parameter range investigated. It might therefore be possible to achieve an even higher etch rate if the parameter range is extended. \\
\indent The third and statistically most significant term is the quadratic dependence on $f_{\textrm{SiCl}_4}$, which reflects the existence of an optimum ratio of SiCl$_4$ to SF$_6$ for achieving a high GaP etch rate (\fref{fig2}(c)). Individual experiments conducted within the scope of the DOE with a low fraction of SiCl$_4$ had a very low etch rate, and processes containing only fluorine species showed virtually no etching, indicating that either SF$_6$ is not a good etchant or it passivates the surface. On the other hand, the etch rate also declines at high SiCl$_4$ content. Indeed, the etch rate for a plasma using only SiCl$_4$ is one and half orders of magnitude lower than the optimum etch rate.  Clearly, species originating from SF$_6$ play a role in accelerating the etching despite the fact that a plasma using SF$_6$ alone hardly etches GaP at all.\\
\indent Over the range of parameters investigated, the model predicts a maximum GaP etch rate of \SI{9900}{nm\per min} at $p_\textrm{C} = \SI{60}{mTorr}$, $V_\textrm{DC} = \SI{250}{V}$, and $f_{\textrm{SiCl}_4} = 73\%$, with the lower bound of the 95\% confidence interval at \SI{3100}{nm\per min} and the upper bound at \SI{32000}{nm\per min}. To test the model’s capability to predict the etch rate with respect to $f_{\textrm{SiCl}_4}$, additional etch experiments were performed on GaP samples at flows of \SI{8}{sccm} (\SI{12}{sccm}), \SI{13}{sccm} (\SI{7}{sccm}), \SI{15}{sccm} (\SI{5}{sccm}) and \SI{20}{sccm} (\SI{0}{sccm}) for SiCl$_4$ (SF$_6$) with $p_\textrm{C} = \SI{60}{mTorr}$, $V_\textrm{DC} = \SI{250}{V}$, and $P_\textrm{ICP} = \SI{300}{W}$. A high ICP power was chosen to ensure the saturation of dissociated and ionized species in the plasma. Although the observed etch rates are somewhat lower than predicted (\fref{fig2}(c)) (presumably because the additional runs were carried out at a later date than the DOE experiments and the environment in the ICP-RIE chamber may be slightly different), the results are within or near the 95\% confidence interval of the fit and follow the shape of the curve nicely. \\
\indent The quality of the model can be summarized in an actual-by-predicted plot, in which the experimentally observed etch rates are compared with those calculated from the model (\fref{fig4}). The coefficient of determination $\textrm{R}^2 = 0.90$ and the probability value $p < 0.0001$ indicate a good fit and a high statistical significance of the model. 

\subsection{Selectivity}
Applying the same DOE analysis method to the selectivity for etching GaP versus Al$_\textrm{x}$Ga$_{1-\textrm{x}}$P for the stoichiometries x = 0.03 and x = 0.10, we again consider only those experiments for which the fraction of SiCl$_4$ is within the range of 37.5\% to 100\% and exclude those samples that exhibited micromasking, namely samples 4, 14, 15, 23, and 24. In this case, we find a considerably more complicated dependence on the various process parameters (\tref{tab3}).  Despite the relatively large number of terms, all but two $p$-values are under 0.001, and the $p$-value for the model as a whole is below 0.0001 for both stoichiometries, indicating quite high statistical significance.

\begin{figure}[t]
	\centering
	\includegraphics[width = 8cm]{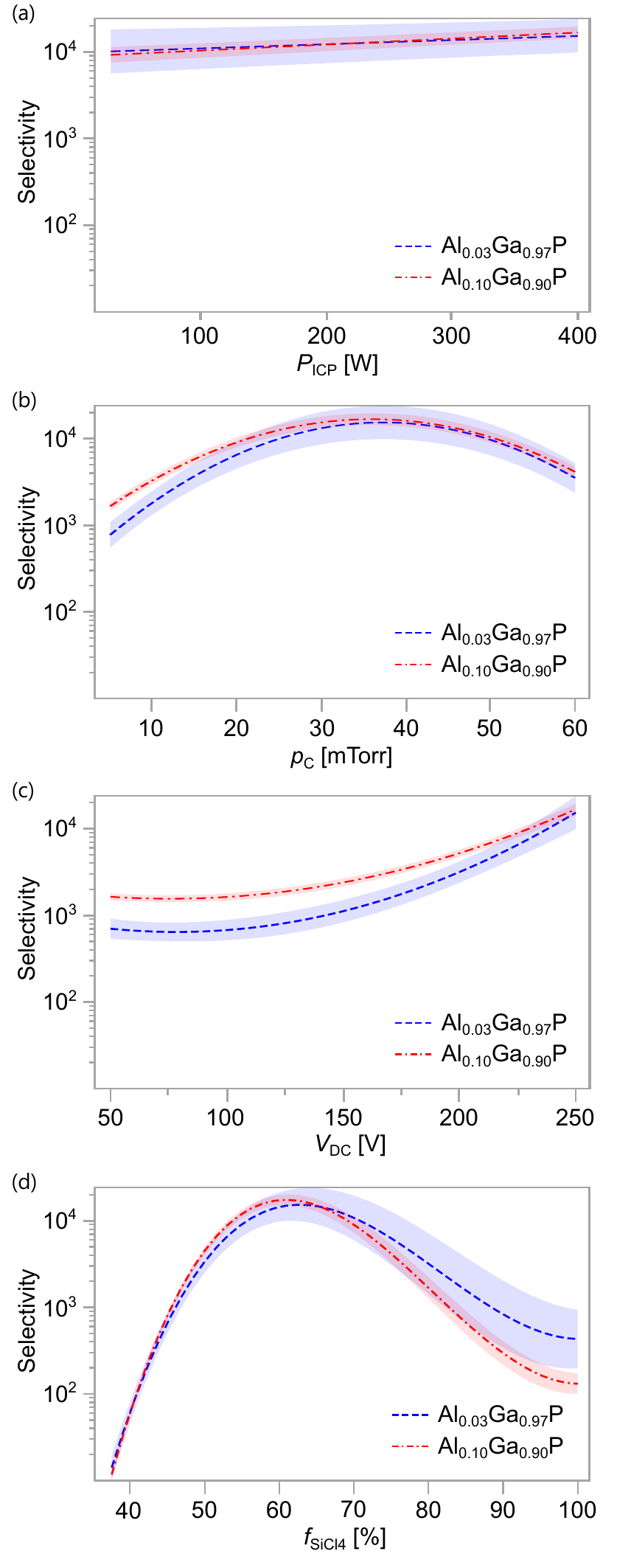}
	\caption{Selectivity as predicted by the model as function of (a) $P_\textrm{ICP}$, (b) $p_\textrm{C}$, (c) $V_\textrm{DC}$ and (d) $f_{\textrm{SiCl}_4}$ (x = 0.03 in blue, x = 0.10 in red). In each plot, the remaining parameters are fixed at $P_\textrm{ICP} = \SI{400}{W}$, $V_\textrm{DC} = \SI{250}{V}$, $p_\textrm{C} = \SI{36.3}{mTorr}$ and $f_{\textrm{SiCl}_4} = 63\%$, respectively, which represents the global maximum of selectivity for x = 0.03. The shaded regions in the respective color indicate the 95\% confidence intervals of the model.}
	\label{fig5}
\end{figure}

\begin{figure}[h]
	\centering
	\includegraphics[width = 8cm]{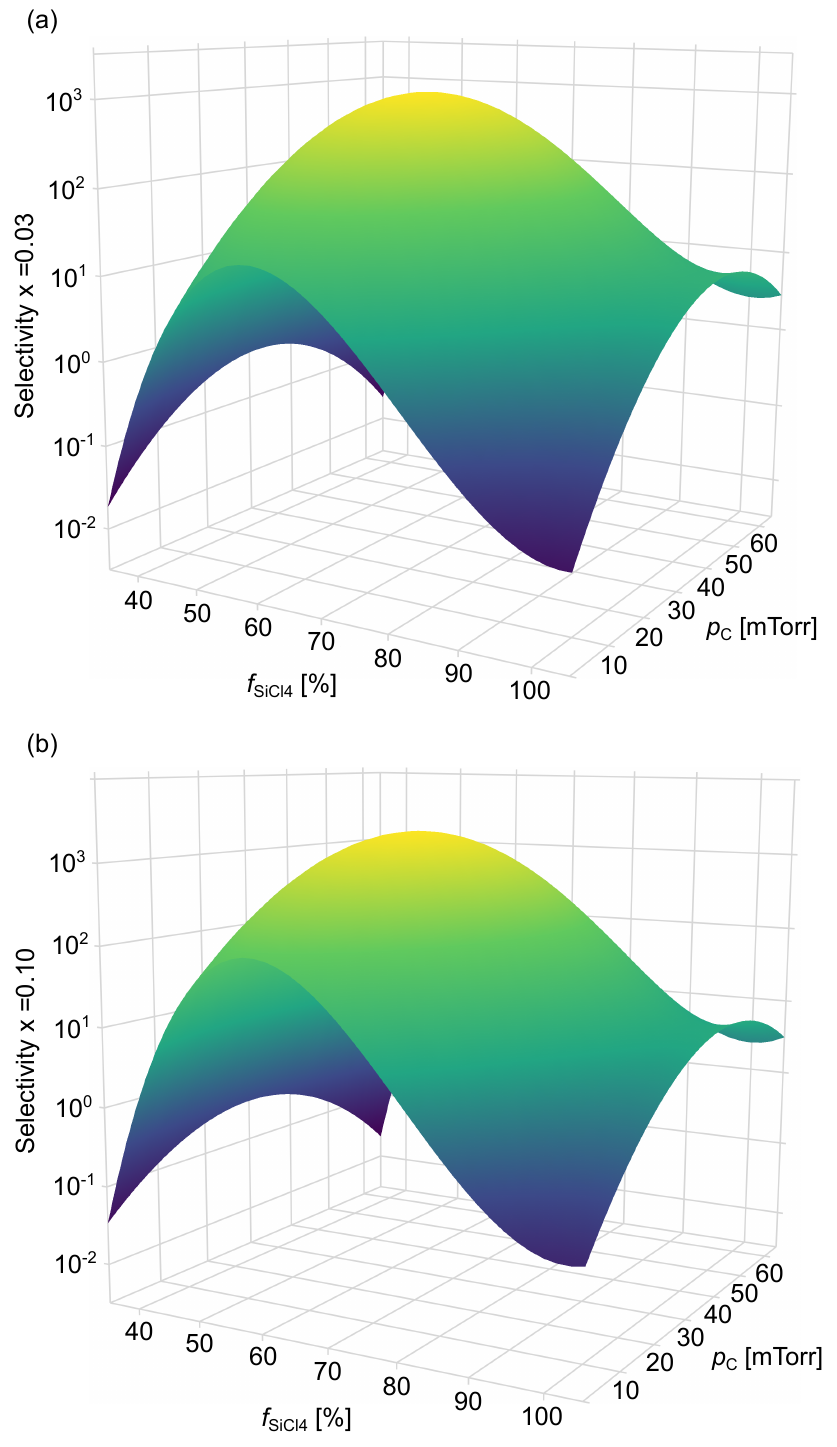}
	\caption{Predicted selectivity for etching GaP over Al$_\textrm{x}$Ga$_{1-\textrm{x}}$P plotted versus chamber pressure ($p_\textrm{C}$) and fraction of SiCl$_4$ ($f_{\textrm{SiCl}_4}$) for (a) x = 0.03 and (b) x = 0.10.}
	\label{fig6}
\end{figure}

\begin{figure}[t]
	\centering
	\includegraphics[width = 8cm]{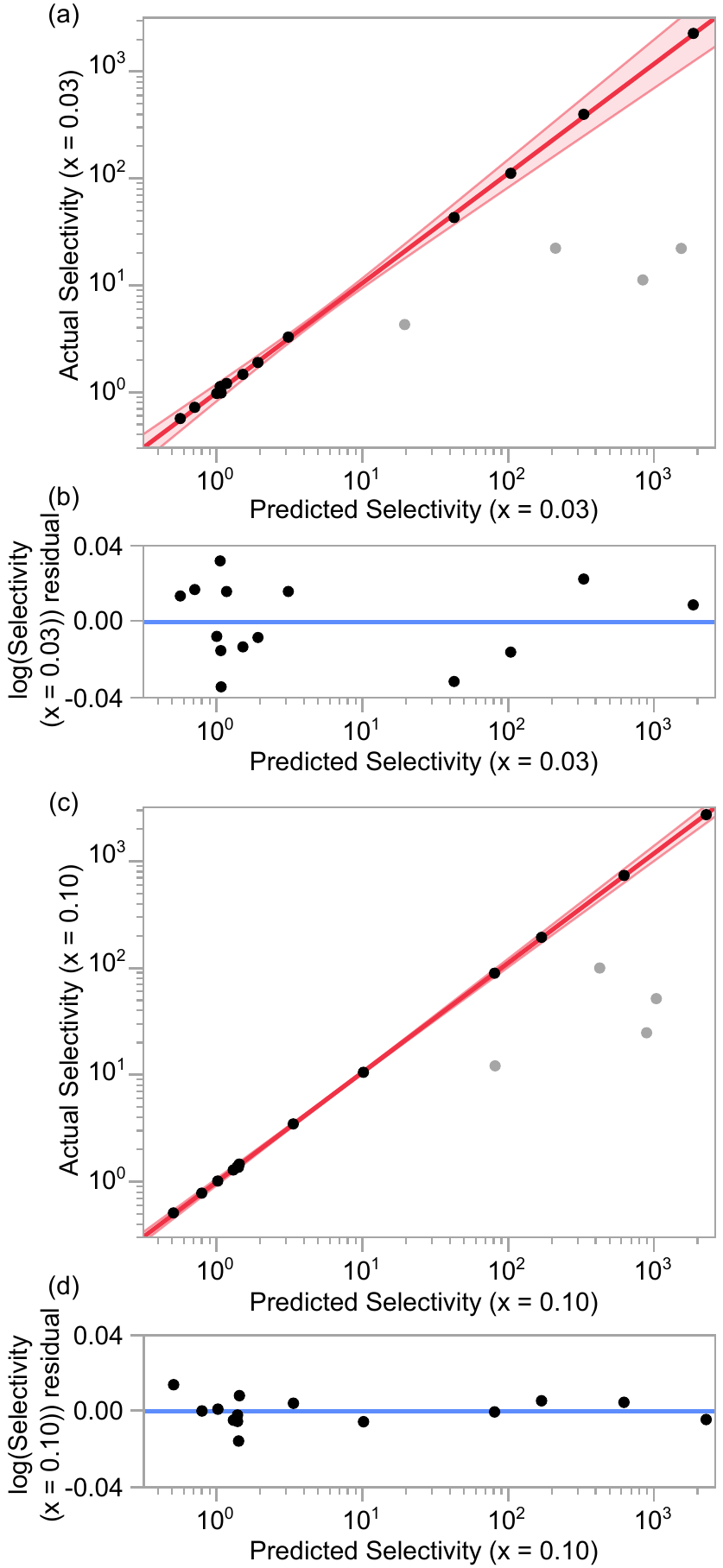}
	\caption{Actual-by-predicted plots of selectivity for etching GaP over Al$_\textrm{x}$Ga$_{1-\textrm{x}}$P for (a) x = 0.03 and (c) x = 0.10, with the 95\% confidence intervals shaded in red, and plots of residuals of the common logarithm of selectivity as a function of predicted selectivity for (b) x = 0.03 and (d) x = 0.10. In the absence of errors, the black dots indicating the individual experiments would lie on the straight lines. The grey dots represent data that were excluded from the analysis due to micromasking on either the GaP or the Al$_\textrm{x}$Ga$_{1-\textrm{x}}$P samples.}
	\label{fig7}
\end{figure}

\begin{table}
	\caption{Model terms and their corresponding probability value $p$ for the selectivity of etching GaP over Al$_\textrm{x}$Ga$_{1-\textrm{x}}$P for x=0.03 and x=0.10.}
	\label{tab3}
	\begin{indented}
		\item[]\begin{tabular}{@{}llll}
			\br
			$\textrm{Term}$ & $p ~(x = 0.03$) & $p ~(x = 0.10)$  \\ \mr
			${f}_{\textrm{SiCl}_4}^2$ & $0.00001$ & $0.00000$ \\ 
			${f}_{\textrm{SiCl}_4}^3$ &$ 0.00035$ & $0.00001$ \\ 
			${f}_{\textrm{SiCl}_4}$ & $0.00062$ & $0.00001$ \\ 
			${V}_{\textrm{DC}}$ & $0.00013$ & $0.00001$ \\ 
			${p}_{\textrm{C}}^2$ & $0.00024$ & $0.00002$ \\ 
			${p}_{\textrm{C}} \cdot \textit{f}_{\textrm{SiCl}_4}$  & $0.00053$ & $0.00003$ \\ 
			${p}_{\textrm{C}}$ & $0.00043$ & $0.00007$ \\ 
			${V}_{\textrm{DC}} \cdot \textit{f}_{\textrm{SiCl}_4}$ & $0.00082$ & $0.00019$ \\ 
			${V}_{\textrm{DC}}^2$ & $0.00224$ & $0.00026$ \\ 
			${P}_{\textrm{ICP}}$ & $0.02276$ & $0.00039$ \\ \br
		\end{tabular}
	\end{indented}
\end{table}
\indent The model predicts a maximum selectivity over the parameter range investigated of roughly 15000:1 for the x = 0.03 stoichiometry at $P_\textrm{ICP} = \SI{400}{W}$, $p_\textrm{C} = \SI{37.3}{mTorr}$, $V_\textrm{DC} = \SI{250}{V}$, and $f_{\textrm{SiCl}_4}$ = 63\%, and 17000:1 for the x = 0.10 stoichiometry at $P_\textrm{ICP} = \SI{400}{W}$, $p_\textrm{C} = \SI{35.6}{mTorr}$, $V_\textrm{DC} = \SI{250}{V}$, and $f_{\textrm{SiCl}_4} = 61\%$, as is evident from \fref{fig5}.  Such selectivity values are beyond the range measurable in our experiments using a \SI{100}{nm}-thick SiO$_2$ hard mask.  \Fref{fig5} also shows the calculated dependence of the selectivity on each individual process parameter with the others fixed at the value that gives the maximum response. The selectivity behavior is qualitatively the same for the two Al$_\textrm{x}$Ga$_{1-\textrm{x}}$P stoichiometries.\\
\indent In contrast to the model for the GaP etch rate, the ICP power is statistically significant in determining the selectivity (\fref{fig5}(a)). Still, it plays the least important role. The modest gain in selectivity with ICP power, which is linear on this logarithmic scale, remains within or close to the 95\% confidence interval.  Again, the ICP power is expected to influence primarily the chemical component of the etching process, and the weak dependence suggests that the plasma composition is already close to being saturated at the lowest value of \SI{25}{W} also for those species involved in the etching of Al$_\textrm{x}$Ga$_{1-\textrm{x}}$P and not just GaP.
\indent The influence of chamber pressure on selectivity (\fref{fig5}(b)) is also different than the situation for the GaP etch rate; several terms containing $p_\textrm{C}$ are statistically significant.  The selectivity varies over an order of magnitude for the higher aluminum content and even more for the lower aluminum content.  In both cases, the selectivity reaches a maximum in the middle of the pressure range evaluated.  Since pressure is statistically insignificant in determining the GaP etch rate, this implies that there is some trade-off governing the Al$_\textrm{x}$Ga$_{1-\textrm{x}}$P etch rate and suggests that, in addition to physical and chemical etching of the surface, other effects such as passivation may play a role. The interdependence of pressure and SiCl$_4$ fraction is illustrated in the response surface plots in \fref{fig6}. \\ 
\clearpage
\indent For DC bias (\fref{fig5}(c)), selectivity appears to be determined by an interplay between the GaP etch rate and the Al$_\textrm{x}$Ga$_{1-\textrm{x}}$P etch rate, both of which are affected by this parameter.  Hence, a quadratic dependence is observed in addition to an overall monotonic increase in selectivity with DC bias. Keeping with the hypothesis of a competition between chemical and physical processes, we attribute the nearly flat selectivity at low DC bias to a relatively weak contribution of the physical component which varies similarly for all stoichiometries.  However, with increasing DC bias, the kinetic energy of the impinging ions accelerates the etching of GaP much more than the etching of Al$_\textrm{x}$Ga$_{1-\textrm{x}}$P.  This behavior is consistent with passivation primarily of Al$_\textrm{x}$Ga$_{1-\textrm{x}}$P.\\
\indent As was the case for the GaP etch rate, the fraction of SiCl$_4$ has the most complex influence on the selectivity (\fref{fig5}(d)).  The dependence is however not just quadratic now but includes linear and cubic terms.  The result is a selectivity that peaks at a lower value of $f_{\textrm{SiCl}_4}$ than the GaP etch rate and that plateaus at high $f_{\textrm{SiCl}_4}$ values where the SF$_6$ flow is low.  This would again be consistent with preferential passivation of Al$_\textrm{x}$Ga$_{1-\textrm{x}}$P, but where the passivation is provided specifically by the SF$_6$.  At high SF$_6$ content, i.e. low $f_{\textrm{SiCl}_4}$ values, both GaP and Al$_\textrm{x}$Ga$_{1-\textrm{x}}$P are etched slowly and selectivity is low because SF$_6$ is a poor etchant even for GaP. With increasing fraction of SiCl$_4$, GaP starts to etch while Al$_\textrm{x}$Ga$_{1-\textrm{x}}$P is passivated. At even higher values of $f_{\textrm{SiCl}_4}$, either the passivation mechanism becomes less effective for Al$_\textrm{x}$Ga$_{1-\textrm{x}}$P or the plasma composition becomes less reactive or both, and selectivity begins declining even before the GaP etch rate peaks. The plateau at the highest values of $f_{\textrm{SiCl}_4}$ is then ascribed to a limiting situation where the differentiation between GaP and Al$_\textrm{x}$Ga$_{1-\textrm{x}}$P is no longer changing due to passivation or plasma reactivity. \\
\indent Overall, the stoichiometry with 10\% aluminum content yields a higher selectivity than the 3\% composition, except in a portion of parameter space where the fraction of SiCl$_4$ is high. As can be seen in \fref{fig5}(d), the selectivity is higher for x = 0.10 at its maximum but drops faster with increasing $f_{\textrm{SiCl}_4}$ than the selectivity for x = 0.03. This suggest that, in addition to the passivation of Al$_\textrm{x}$Ga$_{1-\textrm{x}}$P by SF$_6$, there may be a competing enhancement at higher aluminum content of etching by chlorine species. \\
\indent The actual-by-predicted plots for selectivity (\fref{fig7}) indicate that the model describes the observations well. For both x = 0.03 and x = 0.10, the coefficient of determination is $\textrm{R}^2 = 1.00$ and the probability value is $p < 0.0001$. We attribute the higher value of R$^2$ for selectivity in comparison to that for the GaP etch rate to the fact that the GaP and Al$_\textrm{x}$Ga$_{1-\textrm{x}}$P samples were etched side by side for each parameter set, which may somewhat reduce any scatter due to run-to-run variations. 

\begin{figure}[h]
	\centering
	\includegraphics[width = 8cm]{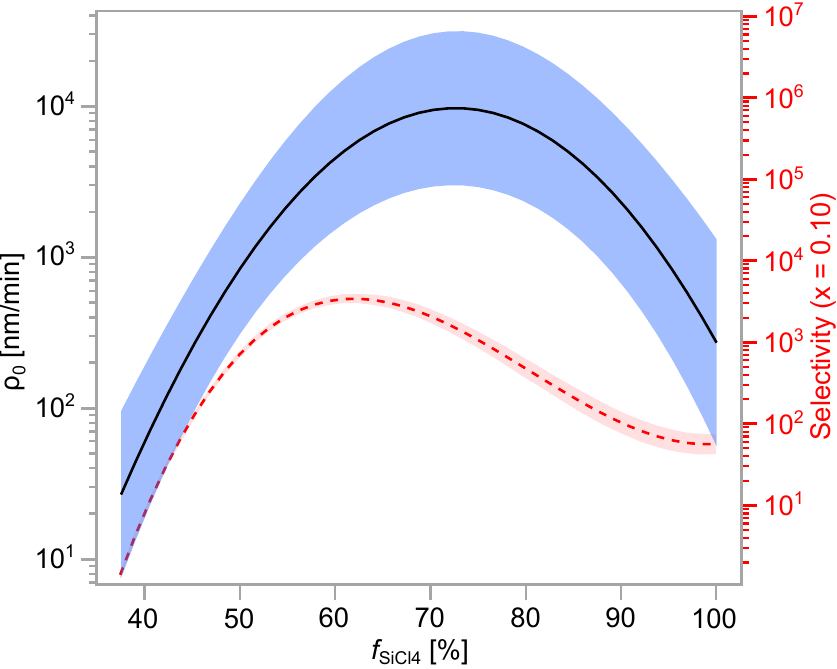}
	\caption{GaP etch rate (black) and selectivity for x = 0.10 (red dashed) as a function of $f_{\textrm{SiCl}_4}$ at $P_\textrm{ICP} = \SI{300}{W}$, $V_\textrm{DC} = \SI{250}{V}$ and $p_\textrm{C} = \SI{60}{mTorr}$. The blue and red shaded regions define the 95\% confidence intervals.}
	\label{fig8}
\end{figure}

\clearpage
\indent  With the set of parameters that gives the maximum selectivity, the model for the GaP etch rate predicts an etch rate of \SI{7900}{nm\per min} with the lower bound of the 95\% confidence interval at \SI{2500}{nm\per min} and the upper bound at \SI{25000}{nm\per min}. On the other hand, the process with the highest predicted GaP etch rate yields a selectivity of 2400:1 with the lower and upper bound of the 95\% confidence interval at 1300:1 and 4300:1, respectively, for x = 0.03.  For x = 0.10, the selectivity is predicted to be 1900:1 with the lower and upper bound of the 95\% confidence interval at 1500:1 and 2300:1, respectively. The process can therefore be adjusted depending on whether high selectivity or high etch rate is required. The two models can be combined in a plot that shows both GaP etch rate and selectivity as a function of $f_{\textrm{SiCl}_4}$ for the specific values of $P_\textrm{ICP} = \SI{300}{W}$, $V_\textrm{DC} = \SI{250}{V}$ and $p_\textrm{C} = \SI{60}{mTorr}$ (\fref{fig8}), from which it is again apparent that the maximum etch rate occurs at a lower fraction of SiCl$_4$ than the maximum selectivity.\\

\clearpage
\indent Finally, we note that even higher selectivities may be attainable with aluminum contents greater than those tested here. From a practical point of view, epitaxial growth of Al$_\textrm{x}$Ga$_{1-\textrm{x}}$P on a GaP substrate with an aluminum content well above 50\% is entirely feasible, as the lattice mismatch between GaP and AlP is only 0.01\% \cite{addamiano1960x}. 

\subsection{Surface Morphology}

\begin{figure}[h]
	\centering
	\includegraphics[width = 8cm]{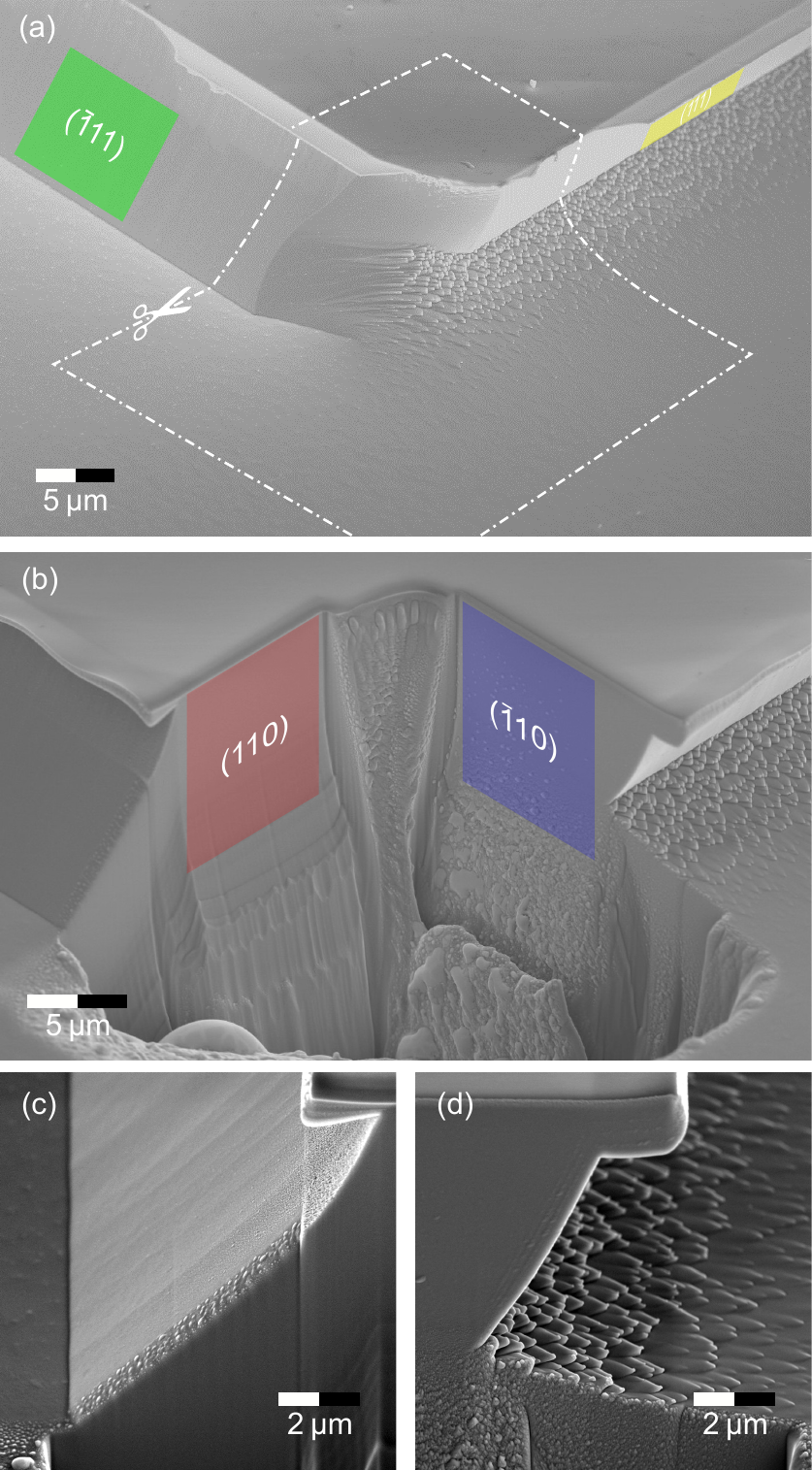}
	\caption{SEM micrographs of an etched square test feature with the GaP crystal planes indicated by color overlays: (a) the as-etched structure with a dashed white line delineating the region subsequently removed by FIB milling; (b) the same structure after FIB milling; (c) and (d) profile views perpendicular to the $(110)$ plane and $(\overline{1}10)$ plane, respectively.}
	\label{fig9}
\end{figure}

The surface morphology of the etched samples varies greatly over the range of process parameters used. Most of the deeply etched GaP samples have a smooth horizontal top surface without any evidence of residues being deposited. Only samples etched in processes 5, 9 and 14 show micromasking to various extents. In both experiment 9 and experiment 14, the fraction of SiCl4 is at the threshold between the passivating domain and the etching domain, so the micromasking may be related to the onset of passivation. The sample for process 5 exhibits micromasking only locally. In contrast, all of the $\textrm{Al}_\textrm{x}\textrm{Ga}_{1-\textrm{x}}\textrm{P}$ samples that were not significantly etched exhibited some residue formation. \\
\indent The etch profile can depend strongly on crystal lattice orientation. \Fref{fig9} shows SEM micrographs of a sample that was etched under conditions for which the GaP etch rate and selectivity are both high ($P_\textrm{ICP} = \SI{303}{W}$, $p_\textrm{C} = \SI{60}{mTorr}$, $V_\textrm{DC} = \SI{130}{V}$, and $f_{\textrm{SiCl}_4} = 60\%$). Here, we examined a square, {100-\SI{}{\micro m}} $\times$ {100-\SI{}{\micro m}}, test feature aligned with the $(110)$ and the $(\overline{1}10)$ flats of the GaP wafer. This allows us to determine the orientation of the crystal lattice planes in the etch profile. \Fref{fig9}(a) displays a micrograph of one corner of the square, whereas figures \ref{fig9}(b) - (d) show the same structure after milling the corner along the dashed white line in \fref{fig9}(a) with a gallium focused ion beam (FIB). The colorized overlays in figures \ref{fig9}(a) and (b) indicate the orientation of the respective lattice planes. The process produces a smooth horizontal top surface as well as smooth, angled and somewhat curved sidewalls which converge to $(\overline{1}11)$ and $(111)$ planes in figures \ref{fig9}(c) and (d), respectively. This behavior is typical of chemical etching of III-V materials, for which etching of the densely packed, group-III-terminated $\{111\}$ planes are often slowest to etch. \cite{gatos1960characteristics,tarui1971preferential}

\section{Discussion}
Literature on etching GaP selectively over Al$_\textrm{x}$Ga$_{1-\textrm{x}}$P is scarce. The only other previously published method is a dry etching process that makes use of SiF$_4$ and SiCl$_4$ and etches GaP at \SI{135}{nm\per min} with a selectivity of 124:1 with respect to Al$_{0.6}$Ga$_{0.4}$P \cite{epple2002dry}. We have observed similar behavior in an ICP-RIE process with CF$_4$ and Cl$_2$ \cite{kscinpress17}.
Etching with a mixture of SiCl$_4$ and SF$_6$ has never before been investigated. However, selective dry-etching of GaAs in the presence of Al$_\textrm{x}$Ga$_{1-\textrm{x}}$As has been repeatedly reported using etching mixtures such as SF$_6$/SiCl$_4$ \cite{salimian1988selective}, CCl$_2$F$_2$  \cite{seaward1987analytical,knoedler1986selective,vatus1986highly,hikosaka1981selective} and SF$_6$/BCl$_3$ \cite{lee2000advanced}. Selectivity in both the GaAs/Al$_\textrm{x}$Ga$_{1-\textrm{x}}$As system and the GaP/Al$_\textrm{x}$Ga$_{1-\textrm{x}}$P system has been attributed to the formation of AlF$_3$ on the surface of the aluminum-containing material which inhibits further etching \cite{epple2002dry,salimian1988selective}. The passivating effect apparently arises from the much lower volatility of AlF$_3$ compared to that of both AlCl$_3$ and GaCl$_3$ (\tref{tab4}). Formation of GaF$_3$, which also has low volatility, is not observed on GaAs or Al$_\textrm{x}$Ga$_{1-\textrm{x}}$As and is presumed to be impeded by a high kinetic barrier to reaction \cite{salimian1988selective,karouta1999high}, explaining why fluorine-containing species provide selectivity but not chlorine species.  In the case of the phosphides, Epple \textit{et al} verified the presence of an aluminum/fluorine-based layer on Al$_{0.6}$Ga$_{0.4}$P by Auger electron spectroscopy after etching with SiF$_4$ and SiCl$_4$ \cite{epple2002dry}. Our models for both the GaP etch rate and the selectivity are fully consistent with the above interpretation.  We see evidence for strong surface passivation of Al$_\textrm{x}$Ga$_{1-\textrm{x}}$P as compared to GaP, where the passivation is associated with SF$_6$. 

\begin{table}
	\caption{Possible reaction products in fluorine- and chlorine-based III-V plasma etch processes and their boiling or sublimation (indicated with $s$) temperature $T_\textrm{b}$ at atmospheric pressure (\SI{101.325}{kPa}). \cite{Lide200306}}
	\label{tab4}
	\begin{indented}
		\item[]\begin{tabular}{@{}ll}
			\br
			$\textrm{Product}$ & $T_\textrm{b} [\SI{}{\degreeCelsius}]$  \\ \mr 
			$\textrm{AlF}_{3}$ & $1276 ~s$ \\ 
			$\textrm{AlCl}_{3}$ & $180 ~s$ \\  
			$\textrm{GaF}_{3}$ & $950 ~s$ \\ 
			$\textrm{GaCl}_{3}$ & $201$ \\  
			$\textrm{NF}_{3}$ & $-128.75$  \\
			$\textrm{PF}_{3}$ & $-101.8$\\ 
			$\textrm{PF}_{5}$ & $-84.6$\\
			$\textrm{PCl}_{3}$ & $76.7$  \\ 
			$\textrm{PCl}_{5}$ & $160 ~s$ \\
			$\textrm{AsF}_{3}$ & $57.8$ \\ 
			$\textrm{AsCl}_{3}$ & $130$ \\\br
		\end{tabular}
	\end{indented}

\end{table}

\indent Although the results of Epple \textit{et al} and those presented here demonstrate that fluorine precursors alone, be it SiF$_4$ or SF$_6$, do not etch GaP, our results indicate that the etching reaction is not simply controlled by the SiCl$_4$; the presence of some SF$_6$ greatly enhances the etch rate (see \fref{fig2}(c)). Similar behavior has been observed both for the GaAs/Al$_\textrm{x}$Ga$_{1-\textrm{x}}$As system \cite{salimian1988selective} and for GaN \cite{karouta1999high} when etching with SF$_6$ and SiCl$_4$, although it is not as dramatic. One contributing factor may be the rate of removal of the group V element, as argued by Karouta \textit{et al} \cite{karouta1999high} for etching of GaN.  The group V halides all have relatively low boiling or sublimation temperatures (\tref{tab4}), but the temperatures for the fluorides are lower.  PF$_3$, PF$_5$, and NF$_3$ are especially volatile, evaporating well below room temperature.  Thus, even though chlorine species may be primarily responsible for etching, fluorine atoms may contribute to the etch process, particularly in the absence of aluminum.  It should be noted though that, if volatility of the reaction products were the only decisive factor, removal of gallium would become rate limiting.  The effect of facile removal of phosphorous must therefore be to lower the reaction barrier to form the gallium containing products.\\
\indent But this explanation alone is not sufficient.  We need to account for our etch rates and selectivity being orders of magnitude higher with a mixture of SF$_6$ and SiCl$_4$ than those previously published for etching with SiF$_4$ and SiCl$_4$. We believe the relative strength with which sulfur and silicon bind fluorine and chlorine may instead be responsible. \Tref{tab5} lists bond dissociation energies for a series of silicon and sulfur halide molecules.  Although we do not know what species are present in the plasma, and there may be a variety of both ions and radicals generated, it is nevertheless clear that fluorine is much more strongly bound to silicon than chlorine, and both halogen atoms are much more weakly bound to sulfur than to silicon.  The use of SF$_6$ instead of SiF$_4$ may thus lead to a transfer of fluorine atoms to silicon but not chlorine atoms to sulfur.  The result would be an increase in reactive chlorine species impinging on the surface to be etched. In fact, during etching of GaAs, Salimian \textit{et al} \cite{salimian1988selective} detected an increase in chlorine atoms in emission spectra when SF$_6$ was added to SiCl$_4$ plasmas.  A related explanation has been proposed to describe etching of GaN with SiCl$_4$/SF$_6$/Ar chemistry \cite{karouta1999high}. 

\begin{table}
	\centering
	\caption{Bond dissociation energies $D^o_{298}$ for a series of silicon and sulfur halide molecules. \cite{Lide200306,walsh1981bond,gailbreath2000potential}}
	\label{tab5}
	\begin{indented}
		\item[]\begin{tabular}{@{}ll}
			\br
			$\textrm{Bond}$ & $D^o_{298} [\SI{}{kJ\per mol}]$  \\ \mr
			$\textrm{F} - \textrm{SiF}_3$ & $669.44$ \\
			$\textrm{F} - \textrm{SF}_5$  & $391.6$ \\
			$\textrm{Cl} - \textrm{SiCl}_3$ & $464.4$ \\ 
			$\textrm{Cl} - \textrm{SF}_5$ & $<272$ \\
			$\textrm{Cl} - \textrm{SCl}$ & $293$ \\\br
		\end{tabular}
	\end{indented}
\end{table}

\indent The net result is that both the rate for etching GaP and the selectivity are strongly dependent on the ratio for SF$_6$ to SiCl$_4$ and the maximum values are reached when both gases are present. We note though that a proper analysis would need to take into account activation energies and the corresponding reaction barriers.

\section{Conclusions}
We successfully demonstrated highly selective etching of GaP in the presence of Al$_\textrm{x}$Ga$_{1-\textrm{x}}$P with a plasma combining SF$_6$ and SiCl$_4$. A design of experiments implemented to model the parameter space yielded a predicted maximum selectivity of 15000:1 or more for etching of GaP over Al$_\textrm{x}$Ga$_{1-\textrm{x}}$P with an Al content as low as 3\% while simultaneously achieving GaP etch rates of several thousand \SI{}{nm/min}. For the parameters tested, selectivities of up to 2700:1 and GaP etch rates above \SI{3000}{nm\per min} were measured experimentally. These results contrast with the previous work using SiF$_4$ and SiCl$_4$ claiming the need for high aluminum content in order to form an etch stop layer and for which etch rates and selectivity were two orders of magnitude lower. Use of a mixture of SiCl$_4$ and SF$_6$ instead of a purely chlorine-based plasma is essential for amplifying the GaP etch rate, which we predict can be tuned to values approaching \SI{10000}{nm\per min}. Although these high-etch-rate and high-selectivity processes exhibit a crystal orientation-dependent morphology, we believe that it may be possible to find less aggressive conditions suitable for pattern transfer applications while maintaining sufficient selectivity. With this advancement in the state of the art of GaP etching, new processing schemes become possible, such as those involving bonding of GaP onto carrier substrates. This opens the door to a variety of new GaP-based integrated nanophotonic applications. 

\section*{Acknowledgments}
This work was supported by funding from the European Union’s Horizon 2020 Programme for Research and Innovation under grant agreement No. 722923 (Marie Curie H2020-ETN OMT) and grant agreement No. 732894 (FET Proactive HOT).
\section*{References}
\bibliographystyle{unsrt}
\bibliography{references}

\end{document}